\documentclass[twocolumn,pra,superscriptaddress,showpacs,amsfonts,amssymb,amsmath,letterpaper,floatfix]{revtex4-1}

\usepackage{amsmath}
\usepackage{graphicx}
\usepackage{amssymb}
\usepackage{pslatex}

\usepackage{color}

\usepackage{txfonts}
\usepackage{dcolumn}
\usepackage{hyperref}

\begin{document}

\title{Transport Signatures of Majorana Quantum Criticality Realized by
Dissipative Resonant Tunneling} 

\author{Huaixiu Zheng}
\affiliation{\textit{Department of Physics, Duke University, P. O. Box 90305,
Durham, North Carolina 27708, USA}}

\author{Serge Florens}
\affiliation{\textit{Institut Neel, CNRS and UJF, 25 avenue des Martyrs, BP 166, 38042 Grenoble, France}}

\author{Harold U. Baranger}
\affiliation{\textit{Department of Physics, Duke University, P. O. Box 90305,
Durham, North Carolina 27708, USA}}

\date{20 March 2014}

\begin{abstract}
We consider theoretically the transport properties of a spinless resonant
electronic level coupled to strongly dissipative leads, in the regime of circuit
impedance near the resistance quantum. Using the Luttinger liquid analogy, one
obtains an effective Hamiltonian expressed in terms of interacting Majorana
fermions, in which all environmental degrees of freedom (leads and
electromagnetic modes) are encapsulated in a single fermionic bath. General
transport equations for this system are then derived in terms of the Majorana
T-matrix. Perturbative treatment of the Majorana interaction term yields the
appearance of a marginal, linear dependence of the conductance on temperature
when the system is tuned to its quantum critical point, in agreement with recent
experimental observations.  We investigate in detail the different crossovers
involved in the problem, and analyze the role of the interaction terms in the
transport scaling functions. In particular, we show that single barrier scaling
applies when the system is slightly tuned away from its Majorana critical point,
strengthening the general picture of dynamical Coulomb blockade.
\end{abstract}


\maketitle

\section{Introduction}

Engineering electronic systems at the nanoscale is becoming a fascinating
way to realize unconventional states of matter, ones that 
break the Fermi liquid paradigm. 
Some recent examples include 
several ways of realizing one-dimensional Luttinger liquid
physics~\cite{ChangRev,DeshpandeRevNature10,ClaessenNP11,LarocheScience14},
gate-tunable molecules showing quantum phase transitions~\cite{RochPRL09,FlorensReview11}, 
and tailored double quantum dots in semiconductors exhibiting 
complex behaviors such as multichannel Kondo physics~\cite{GoldhaberGRev07A,Potok2CK07}. 
Further progress and new classes of anomalous behavior can be realized by
combining {\it both} fine-tuned nanostructures and tailored environments,
as demonstrated by a series of recent 
experiments~\cite{ParmentierPierreNatPhys11,MebrahtuNat12,MebrahtuNatPhys13,JezouinPierreNatComm13}
involving quantum tunneling at the nanoscale in the presence of strong dissipation in the 
contacts. In this type of system, single electron tunneling events
create large electromagnetic fluctuations, that become energetically
prohibitive in a strongly resistive circuit. This so-called dynamical Coulomb
blockade phenomenon leads to inelastic losses that can be
quite effective in impeding low-energy electrons from transporting current, and
so dramatically depress the conductance for small applied voltage bias 
across the device (typically in a power-law fashion).
This physical behavior is quite reminiscent of the problem of quantum tunneling
in Luttinger liquids, one-dimensional conducting wires where Coulomb interaction
effects are prominent~\cite{GiamarchiBook}. 
In that case, power-law zero-bias anomalies in transport
also arise due to excitations of collective plasmon modes. This analogy can
be formally pushed to a general theoretical equivalence between the two problems 
using bosonization techniques~\cite{SafiSaleurPRL04}, which makes dissipative
circuits an attractive method for probing local aspects of Luttinger liquid physics in
nanocircuits.

Recent experimental investigations further explored this
analogy by extending previous single barrier devices to quantum dot
systems~\cite{MebrahtuNat12,MebrahtuNatPhys13}. 
Here, additional quantum degrees of freedom are introduced, such
as the quantized charge and magnetic moment for the localized electronic
level. Previous theoretical arguments~\cite{MebrahtuNat12,FlorensPRB07}
showed that the Luttinger analogy is still maintained, opening an
interesting playground for quantum critical and anomalous Kondo-type 
behavior. 

In the present paper, we aim at analyzing in detail the
transport characteristics in the simpler case when only the local electron 
charge is the relevant variable, as can be realized by a full spin-polarization
of the electronic states in a large magnetic field. This situation 
results in complex signatures because zero bias anomalies
are very sensitive to the typical transmission through the device. They can, for
instance, be washed out when the transmission of the electron channel approaches
unity.
While the complete loss of dynamical Coulomb blockade at perfect transmission is 
correct for single tunnel barriers, it turns out to be quite non-trivial in the 
case of resonant tunneling through a perfectly transmitting electronic level. 

We show here that full transmission 
does survive large dissipation in the contacts, but extra energy loss in the
environment is still possible which then modifies the low-temperature behavior of
the conductance. This behavior can be rationalized 
when the dissipation is fine-tuned such that the
impedance is close to the quantum value $h/e^2$
(here $h$ is Planck's constant and $e$ the electron charge), 
where an exact mapping to resonant Majorana levels can be achieved at low energy.
Losses in the circuit are then embodied in Majorana interaction terms, that were
discarded in previous theoretical studies~\cite{KomnikPRL03}.
In contrast, we show that these terms are not only large
in magnitude for 
dissipative circuits, but even control the leading behavior of the conductance 
near the unitary limit. Here, a striking behavior of the inelastic scattering 
rate---\emph{linear} in temperature and voltage---is obtained, which we view as a 
hallmark of interacting Majorana quantum criticality that was uncovered in recent 
experimental studies~\cite{MebrahtuNatPhys13}. 

A further question that we wish to examine here is to what extent single barrier 
scaling applies to the quantum dot setup when the system deviates from the 
resonance condition. We show that corrections due to the Majorana interaction term 
are in this case---and in contrast to the resonant case mentioned above---very 
rapidly suppressed at low temperature, typically as $T^4$. This result vindicates 
the use of usual dynamical Coulomb blockade theory in a more general way than 
previously thought. 

The paper is organized as follows. In Sec.~\ref{Model} we present our
model of resonant tunneling with dissipation, and outline the connection
to Luttinger and Majorana physics. In Sec.~\ref{Transport}, we present
a general theory of transport formulated in the Majorana language and
provide a perturbative treatment of inelastic processes, leading to a 
detailed study of various transport scaling laws in Sec.~\ref{Scaling}.

\section{Majorana representation of dissipative resonant tunneling}
\label{Model}

\subsection{Modeling a resonant level with dissipative leads}

We present here the basic model for resonant tunneling through a single spin-polarized 
electronic level with resistive leads characterized by the dimensionless 
quantity $r=Re^2/h$, 
the ratio of the lead zero-frequency impedance $R$ to the resistance quantum $h/e^2$.  
For simplicity, we drop spin indices. Our starting Hamiltonian reads
\begin{equation}
H=H_{\textrm{dot}}+H_{\textrm{leads}}+H_{\textrm{T}}+H_{\textrm{env}},
\label{eq:H}
\end{equation}
where $H_{\textrm{dot}}=\epsilon_{\textrm{d}}d^{\dagger}d$ is the Hamiltonian
representing the dot with a single energy level $\epsilon_{\textrm{d}}$ (tuned by the 
backgate voltage $V_\mathrm{gate}$), and 
$H_{\textrm{leads}}=\sum_{\textrm{\textrm{\ensuremath{\alpha}=S,D}}} 
\sum_{k}\epsilon_{k}c_{k\alpha}^{\dagger}c_{k\alpha}$
describes the electrons in the source (S) and drain (D) electrodes.
Tunneling between the dot and the leads with amplitudes $V_{S/D}$ is
given by
\begin{equation}
H_{\textrm{T}}=V_{S}\sum_{k}(c_{kS}^{\dagger}e^{-i\varphi_{S}}d+  {\rm h.c.} )
+V_{D}\sum_{k}(c_{kD}^{\dagger}e^{i\varphi_{D}}d+ {\rm h.c.} ),
\label{eq:Htunnel}
\end{equation}
where the operators $\varphi_{S/D}$ describe phase fluctuations of the
tunneling amplitude between the dot and the S/D lead. These phase operators 
are canonically conjugate to the charge operators $Q_{S/D}$ associated with
the S/D junctions. 
Here, we have adopted the standard treatment quantum tunneling in the presence of a 
dissipative environment \cite{IngoldNazarov92}, which is valid for electrons propagating 
much slower than the electromagnetic field \cite{Nazarovbook}. 

It is useful to transform to phase variables related to the total charge on the dot.
To that end, we introduce \cite{IngoldNazarov92} two new phase operators,
\begin{eqnarray}
\varphi_{S} & \equiv & \kappa_{S}\varphi+\psi \qquad\nonumber \\
\varphi_{D} & \equiv & \kappa_{D}\varphi-\psi \;,\label{eq:}
\end{eqnarray}
where $\kappa_{S/D}=C_{S/D}/(C_{S}+C_{D})$ in terms of the capacitances of the
dot to the source/drain contacts, $C_{S/D}$. The phase $\psi$ is the variable 
conjugate to the fluctuations of total charge on the dot $Q_{c}=Q_{S}-Q_{D}$ and 
so couples to voltage fluctuations on the gate which controls the energy level 
of the dot. 
Likewise, $\varphi$ is the variable conjugate to the charge transferred accross
the device, $Q=(C_{S}Q_{D}+C_{D}Q_{S})/(C_{D}+C_{S})$. Assuming for simplicity 
$C_{S}=C_{D}$, we have $\varphi_{S}=\varphi/2+\psi$ and $\varphi_{D}=\varphi/2-\psi$. 

The gate voltage fluctuations will be disregarded here, as the gate capacitance
in the experiment of Ref.~\cite{MebrahtuNat12,MebrahtuNatPhys13} was negligible, 
$C_{g}\ll C_{S/D}$. 
(The opposite limit of a strongly fluctuating gate coupled to a resonant level but with no dissipation in the leads was considered theoretically in Refs.\,\cite{ImamAverinPRB94,ButtikerPRL00,LeHurPRL04,LeHurLiPRB05,BordaPRB05,BordaZarandX06,GlossopIngersent07,ChungQPT,ChengIngersent09}, and the combination of both types of dissipation was recently treated in Ref.\,\cite{Dong2bathsPRB14}.) 
Thus, only the {\it relative} phase difference between the two leads remains
\cite{IngoldNazarov92,FlorensPRB07}, and the tunneling Hamiltonian becomes
\begin{equation}
H_{\textrm{T}}=V_{S}\sum_{k}(c_{kS}^{\dagger}e^{-i\frac{\varphi}{2}}d
+ {\rm h.c.})+V_{D}\sum_{k}(c_{kD}^{\dagger}e^{i\frac{\varphi}{2}}d+ {\rm h.c.} ). 
\end{equation}

The last part of Eq. (\ref{eq:H}) is the Hamiltonian of the environment,
$H_{\textrm{env}}$ \cite{CaldeiraPRL81,LeggettRMP87,IngoldNazarov92}. The environmental
modes are represented by harmonic oscillators controlled by inductances and
capacitances such that the frequency of environmental modes are given by
$\omega_{k}=1/\sqrt{L_{k}C_{k}}$. These oscillators are then bilinearly coupled
to the phase operator $\varphi$ through the relevant phase variable:
\begin{equation}
H_{\textrm{env}}=\frac{Q^{2}}{2C}+\sum_{k=1}^{N}\left[\frac{q_{k}^{2}}{2C_{k}}
+\left(\frac{\hbar}{e}\right)^{2}\frac{1}{2L_{k}}\left(\varphi-\varphi_{k}\right)^{2}\right]. 
\end{equation}

\subsection{The Luttinger bosonic representation}
Now, we use bosonization~\cite{GiamarchiBook} to map model (\ref{eq:H}) to the
Hamiltonian of a resonant level contacted to two Luttinger liquids. 
Here, we follow closely previous work on tunneling through a single barrier with an 
environment \cite{SafiSaleurPRL04,LeHurLiPRB05} and the Kondo effect in the presence of resistive leads \cite{FlorensPRB07} (see also our previous work in Refs.\,\cite{MebrahtuNat12,MebrahtuNatPhys13}).
The source and drain leads can be standardly reduced to two semi-infinite non-chiral 
one-dimensional free fermionic baths~\cite{GiamarchiBook}.
By an unfolding procedure, one obtains two infinitely-propagating chiral 
fields~\cite{GiamarchiBook}, which both couple to the dot at the origin $x=0$.
One can then bosonize the fermionic fields~\cite{GiamarchiBook} as
$c_{S/D}(x)=\frac{1}{\sqrt{2\pi a_0}} \exp[i\phi_{S/D}(x)]$ (we neglect Klein factors 
for simplicity as their role is unimportant here), where $\phi_{S/D}$ are the bosonic fields introduced to
describe the electronic states in the leads, and $a_0$ is a short distance cutoff. 
Defining the flavor field $\phi_{f}$ and charge field $\phi_{c}$ by
\begin{equation}
\phi_{f} \equiv \frac{\phi_{S}-\phi_{D}}{\sqrt{2}}, \qquad
\phi_{c} \equiv \frac{\phi_{S}+\phi_{D}}{\sqrt{2}},
\label{eq:defphifc}
\end{equation}
one can rewrite the lead Hamiltonian as
\begin{equation}
H_{\textrm{leads}}=\frac{v_{F}}{4\pi}\int_{-\infty}^{\infty}dx
\left[\left(\partial_{x}\phi_{c}\right)^{2}+\left(\partial_{x}\phi_{f}\right)^{2}\right]. 
\end{equation}
with $v_F$ the Fermi velocity. The tunneling Hamiltonian then becomes
\begin{eqnarray}
H_{\textrm{T}} & = & \frac{V_{S}}{\sqrt{2\pi a_0}}\,
\text{exp}\left[-i\frac{\phi_{c}(0)+\phi_{f}(0)}{\sqrt{2}}-i\frac{\varphi}{2}
\right] d+ {\rm h.c.} \nonumber \\
&  + & \frac{V_{D}}{\sqrt{2\pi a_0}}\,
\text{exp}\left[-i\frac{\phi_{c}(0)-\phi_{f}(0)}{\sqrt{2}}
+i\frac{\varphi}{2}\right] d+ {\rm h.c.}.
\end{eqnarray}

A key feature of $H_{\textrm{T}}$ is that the fields $\varphi$ and $\phi_{f}(0)$ enter in the same way in the tunneling process. Combining these two fields together embodies a local tunneling process which is analogous to having effectively interacting leads as in a Luttinger liquid. We thus combine the phase factors as
\begin{subequations}
\label{eq:defphifprime}
\begin{eqnarray}
\phi_{f}' &\equiv&
\sqrt{g}\left(\phi_{f}(0) +\frac{1}{\sqrt{2}}\varphi\right),\qquad \\
\varphi' &\equiv& \sqrt{g}\left(\sqrt{r}\phi_{f}(0)
-\frac{1}{\sqrt{2r}}\varphi\right),
\end{eqnarray}
\end{subequations}
where $g\equiv 1/(1+r)\leq1$ and the new fields are scaled so that they are free fields away from the tunneling points.
The action describing the tunneling in terms of the new phase variables then reads
\begin{eqnarray}
\nonumber
S_{\textrm{T}}&=&\int d\tau \Bigg[\frac{V_{S}}{\sqrt{2\pi a_0}}
e^{-i\frac{1}{\sqrt{2}}\phi_{c}(\tau)}e^{-i\frac{1}{\sqrt{2g}}\phi_{f}'(\tau)}d
+{\rm c.c.}\\
&&
+\frac{V_{D}}{\sqrt{2\pi a_0}}
e^{-i\frac{1}{\sqrt{2}}\phi_{c}(\tau)}e^{i\frac{1}{\sqrt{2g}}\phi_{f}'(\tau)}d
+ {\rm c.c.}\Bigg] \;. 
\label{eq:LT}
\end{eqnarray}

Because of the local nature of the tunneling Hamiltonian, one can proceed with an 
integration over all phase modes away from the origin as well as of the environmental modes. This leads to an effective action for the combined leads and environment given by~\cite{KanePRB92,FurusakiPRB93,LeggettRMP87,SafiSaleurPRL04,FlorensPRB07}
\begin{equation}
S_{\textrm{Leads+Env}}^{\textrm{eff}}\!\!=\frac{1}{\beta}\sum_{n}|\omega_{n}|
\left(|\phi_{c}(\omega_{n})|^{2}+|\phi_{f}'(\omega_{n})|^{2}+|\varphi'(\omega_{n})|^{2}\right) ,
\label{eq:Seff}
\end{equation}
with $\omega_{n}=2\pi n T$ a Matsubara frequency ($T$ is temperature,
$\beta=1/T$, and $n$ is an integer). 
It turns out that one obtains a very similar effective action 
by starting from a model of spinless resonant level
coupled to Luttinger liquids \cite{KanePRB92,FurusakiPRB93,EggertAffleckPRB92},
with Luttinger parameter $g$ ($g<1$ for repulsive interactions). Thus, in
the absence of dissipation, $r=0$, one recovers the correct limit of 
non-interacting fermions $g=1$.

\subsection{The Majorana mapping}
\label{sec:MajoranaMap}

In this last step, we concentrate on the special value $r=1$, corresponding to a 
fine-tuned circuit impedance $R=h/e^2$ (close to the experimental value of
Ref.\,\cite{MebrahtuNatPhys13}), which admits making interesting analytical progress.
We use here the refermionization~\cite{GiamarchiBook} of the tunneling 
term~(\ref{eq:LT}), which starts by performing a unitary transformation
\cite{EmeryKivelsonPRB92,KomnikPRL03}, $U=\exp[i(d^{\dagger}d-1/2) \phi_c(0) / \sqrt{2}]$, 
in order to eliminate the $\phi_c$ charge field in the tunneling action, 
Eq. (\ref{eq:LT}):
\begin{equation}
S_{\textrm{T}} =
\int \!\!d\tau \Big[\frac{V_{S}}{\sqrt{2\pi a_0}}
e^{-i\frac{1}{\sqrt{2g}}\phi_{f}'(\tau)}d
+\frac{V_{D}}{\sqrt{2\pi a_0}}
e^{i\frac{1}{\sqrt{2g}}\phi_{f}'(\tau)}d \Big]
+ {\rm c.c.}\;. 
\label{eq:ST}
\end{equation}
This operation generates a new contact interaction between the dot and the 
phase field:
\begin{equation}
 H_{C}= -\pi v_F \,(d^{\dagger}d-1/2)\, \partial_x \phi_c(x=0) \;.
\end{equation}
For the special value $g=1/2$, corresponding to $r=1$, one can identify
fictitious but emergent fermionic fields $\psi_{c}=e^{i\phi_{c}}/\sqrt{2\pi a_0}$ 
and $\psi_{f}=e^{i\phi_{f}'}/\sqrt{2\pi a_0}$.
Electron waves in the contacts and environment fluctuations in the circuit are thus 
combined together in a non-trivial way into non-interacting (free) fermionic species. 
All the complexity of the tunneling process now reduces to the form
\begin{subequations}
\begin{eqnarray}
\label{eq:DefHMajorana}
H_\textrm{Majorana}&\equiv&
H_T + H_\textrm{dot}+ H_C \\
\label{eq:Hfinal}
&=& \big[ V_S \,\psi_f^\dagger(0) \, d +{\rm h.c.}\big] + \big[V_D \,\psi_f(0) \, d +{\rm h.c.}\big]\\
\nonumber
&& + \, \epsilon_d \, d^\dagger d  -\pi v_F \,\colon\psi_c^\dagger(0) \psi_c(0)
\colon\, 
(d^\dagger d - 1/2),
\end{eqnarray}
\end{subequations}
where $H_\textrm{Majorana}$ describes everything not included in the harmonic leads and environment, 
$H = H_\textrm{leads} + H_\textrm{env} + H_\textrm{Majorana}$. 
A remarkable feature of this effective Hamiltonian is the presence of ``pairing''
terms, like $\psi_f(0) d$, in contrast to the initial tunneling Hamiltonian
Eq.\,(\ref{eq:Htunnel}) where the number of fermions is conserved. 
The underlying reason for the appearance of these pairing terms is that current
in the source-drain circuit is produced both by destroying an electron on the
dot while moving it to the drain and by moving an electron from the source to
the dot; hence, $\psi_f$ (the field describing the current) couples to both $d$
and $d^\dagger$.
This structure motivates the introduction of a Majorana description of 
the local electronic level, 
\begin{equation}
\gamma_1 \equiv \frac{d+d^\dagger}{\sqrt{2}} \quad \text{and} \quad 
\gamma_2 \equiv \frac{d-d^\dagger}{\sqrt{2}i},
\end{equation}
so that $\gamma_1$ and $\gamma_2$ obey
$\gamma_1^\dagger=\gamma_1$, $\gamma_2^\dagger=\gamma_2$,
$\{\gamma_1,\gamma_2\}=0$, and $\gamma_1^2=\gamma_2^2=1/2$.
The effective tunneling Hamiltonian~(\ref{eq:Hfinal}) then becomes
\begin{eqnarray}
\label{eq:HMajorana}
H_\textrm{Majorana}&=&
(V_S-V_D)\frac{\psi_f^\dagger(0)-\psi_f(0)}{\sqrt{2}} \gamma_1 \\ 
\nonumber
&& + i (V_S+V_D) \frac{\psi_f^\dagger(0)+ \psi_f(0)}{\sqrt{2}} \gamma_2 \\
\nonumber
&& + i \epsilon_d \,\gamma_1 \gamma_2  + i \lambda
\,\colon\psi_c^\dagger(0)\psi_c(0) \colon
\,\gamma_1 \gamma_2,
\end{eqnarray}
with $\lambda=- \pi v_F$.

A very special working point can be identified from Hamiltonian~(\ref{eq:HMajorana}): 
$V_S=V_D$ and $\epsilon_d=0$ corresponding to symmetric tunneling amplitudes to source and drain and exactly on resonance. In that case the $\gamma_1$ Majorana mode does
not hybridize to either the leads or the $\gamma_2$ Majorana level; the latter
is, however, tunnel coupled to the fermion bath. If one momentarily forgets
the contact interaction [last term in
Eq.~(\ref{eq:HMajorana})], one obtains the solvable Emery-Kivelson 
point~\cite{EmeryKivelsonPRB92,KomnikPRL03}, described by a non-interacting Majorana
resonant level model for mode $\gamma_2$ together with a perfectly decoupled
Majorana mode $\gamma_1$. This leads to a Majorana quantum critical state with fractional 
degeneracy (the ground state entropy is then $S=\log[\sqrt{2}$]).
In our case, the interaction strength $\lambda$ is, however, large and certainly cannot
be neglected. 
\emph{One purpose of the present paper is to investigate the
consequences of this contact interaction---we will see that it strongly affects
the quantum critical properties.} 

We note finally that for $r$ close to one,
one obtains a Majorana model equivalent to Eq.~(\ref{eq:HMajorana}), but now with 
weakly interacting Luttinger fermionic
fields~\cite{FabrizioPRB95,GoldsteinPRB10,AffleckPrivCom}, described by a new effective 
Luttinger parameter $\tilde{g}-1 \approx (1-r)/2$.
This residual interaction among the fermions leads to slight modifications
of the transport laws derived in the following, but without affecting dramatically, we believe, the general picture.  Although the critical state is then not exactly described
by a Majorana zero mode, the associated ground state still possesses entropy 
$S=\log[\sqrt{1+r}]$  associated with a non-trivial fractional 
degeneracy~\cite{WongAffleck94}.

\section{General transport theory of interacting Majorana modes}
\label{Transport}

We now investigate in detail the conductance through the dot for $r=1$, both at
and away from the critical state, taking into account the Majorana interaction
term.  It is natural to split the Majorana Hamiltonian~(\ref{eq:HMajorana}) into
non-interacting and interacting parts, $H_\textrm{Majorana}=H_0+H_C$, allowing a
perturbative treatment of $H_C$. We are guided by similar perturbative
treatments near the Emery-Kivelson point in other physical systems in which
thermodynamic quantities as well as the bulk resistivity have been
calculated~\cite{SenguptaPRB94,ColemanIoffeTsvelik,ZitkoPRB11}.  A general
conductance formula is first derived in the Majorana description, and then it is
evaluated perturbatively to second order. 

\subsection{Current operator in Majorana terms}

The starting point for the derivation of a general conductance formula is the
current operator, $I \equiv i \big[(N_S - N_D)/2,\, H\big]$ where $N_{S/D}$
denote the number operators for the original fermions in the leads. Applying the
transformations in Eqs.\,(\ref{eq:defphifc}) and (\ref{eq:defphifprime}) and
noting that the unitary operator applied in Sec.\,\ref{sec:MajoranaMap} does not
affect the current operator~\cite{Dong2bathsPRB14}, we find 
\begin{equation}
 I = \frac{i}{2}\left[N_f,H \right] =
 \frac{i}{2}\left\{ V_S\psi^{\dagger}_f(0)-V_D\psi_f(0)\right\} d+{\rm h.c.},
\label{eq:Current}
\end{equation}
using the refermionized form of the tunneling amplitude, Eq.~(\ref{eq:HMajorana}), and denoting the number operator for the transformed $\psi_f$ fermions 
by $N_f$.

In the rest of this paper, we focus on the \emph{symmetric} coupling case,
$V_S=V_D \equiv V$, and examine scaling laws both in the vicinity of and away
from the Majorana quantum critical point by \emph{tuning} the level position 
$\epsilon_{\textrm{d}}$.  
It turns out to be advantageous to introduce a Majorana fermion representation for the
fermionic bath $\psi_f$ as well:
\begin{eqnarray}
 a(x)&\equiv&\frac{\psi_f(x)+\psi^{\dagger}_f(x)}{\sqrt{2}},\qquad
b(x)\equiv\frac{\psi_f(x)-\psi^{\dagger}_f(x)}{\sqrt{2}i}.
\label{eq:Majorana_Rep}
\end{eqnarray}
The tunneling Hamiltonian and contact interaction appearing in Eq.~(\ref{eq:HMajorana})
can then be rewritten as
\begin{equation}
 H_T=2i\,V \,a(0) \,\gamma_2,\qquad H_C=i\lambda \,\gamma_1 \,\gamma_2 
 \,\colon\!\! \psi^{\dagger}_c(0)\psi_c(0)\colon,
\label{eq:HTHC_MR}
\end{equation}
and the current operator becomes simply 
\begin{equation}
I=i\sqrt{2}\, V \,b(0)\,\gamma_2 .
\end{equation}

\subsection{Majorana Green functions}

We wish to find the linear response conductance \cite{BruusBook}
\begin{equation}
 G=-\underset{\omega\rightarrow0}{\text{lim}}\;
\frac{e^2}{\hbar\omega}\text{Im}C^{R}_{II}(\omega),
\end{equation}
where the retarded current-current correlator can be obtained via the analytic
continuation of the Matsubara frequency correaltor,
$C^{R}_{II}(\omega)=C_{II}(i\omega_n\rightarrow \omega+i\eta)$. The Matsubara
correlator $C_{II}(i\omega_n)$ is in turn given by~\cite{BruusBook}
\begin{subequations}
\label{eq:CII}
\begin{eqnarray}
 C_{II}(i\omega_n)&=&\int^{\beta}_0 d\tau e^{i\omega_n\tau}C_{II}(\tau), \\
 C_{II}(\tau)&=&-\langle T_{\tau}I(\tau)I(0)\rangle
=-\frac{\text{Tr}[e^{-\beta H} T_{\tau} I(\tau)I(0)]}{\text{Tr}[e^{-\beta H}]},
\end{eqnarray}
\end{subequations}
where $T_{\tau}$ is the time ordering operator in imaginary time. 
$C_{II}(i\omega_n)$
can be computed using the Matsubara frequency Green function 
method, with the basic non-interacting Green functions of Majorana fermions defined as
\begin{equation}
 G^{(0)}_{AB}(\tau)\equiv-\langle T_{\tau} A(\tau)B(0)\rangle_0
=-\frac{\text{Tr}[e^{-\beta H_0} T_{\tau} A(\tau)B(0)]}{\text{Tr}[e^{-\beta H_0}]}, 
\label{eq:GAB}
\end{equation}
where $A$, $B=a(0)$, $b(0)$, $\gamma_1$ or $\gamma_2$. 
Notice that Eqs.\,(\ref{eq:CII}b) and (\ref{eq:GAB}) are evaluated under, respectively, the full Hamiltonian $H_\textrm{Majorana}$ and the non-interacting Hamiltonian $H_0$ .

Using the equation of motion technique \cite{BruusBook}, one readily finds the
non-interacting ($\lambda=0$) Green functions exactly. The retarded free Green functions in
frequency space are
\begin{subequations}
\label{eq:GF_0}
\begin{eqnarray}
&& \!\!\!\!\!\! \left(\begin{array}{cc}
G^{(0)}_{\gamma_1\gamma_1}(\omega) & G^{(0)}_{\gamma_1\gamma_2}(\omega) \\[2mm]
G^{(0)}_{\gamma_2\gamma_1}(\omega) & G^{(0)}_{\gamma_2\gamma_2}(\omega) \\
\end{array}
\right)=\frac{1}{\omega(\omega+i\Gamma)-\epsilon_{\textrm{d}}^2}
\left(\begin{array}{cc}
\omega+i\Gamma & i\epsilon_{\textrm{d}} \\
-i\epsilon_{\textrm{d}} & \omega \\
\end{array}
\right) \quad\;\; \label{eq:G0gamgam}\\[2mm]
&& G^{(0)}_{a(0)a(0)}(\omega)=
-i\pi\rho\left[1+\frac{-i\Gamma\omega}{\omega(\omega+i\Gamma)-\epsilon_{\textrm{d}}^2}\right] \\[2mm]
&& \left( \begin{array}{c}
G^{(0)}_{a(0)\gamma_1}(\omega) \\[2mm]
G^{(0)}_{a(0)\gamma_2}(\omega) \\
\end{array}
\right)= \frac{-2i\pi\rho V}{\omega(\omega+i\Gamma)-\epsilon_{\textrm{d}}^2}
\left(\begin{array}{c}
 \epsilon_{\textrm{d}} \\
i\omega \\
\end{array}
\right) \label{eq:G0agam}\\[2mm]
&& G^{(0)}_{b(0)b(0)}(\omega)=-i\pi\rho,\\[2mm]
&& G^{(0)}_{b(0)A}(\omega)
=0, \quad A=a(0),\; \gamma_{1},\; \text{or}\; \gamma_{2},
\end{eqnarray}
\end{subequations}
where $\Gamma=4\pi\rho V^2$ and $\rho$ is the electronic density of states. From 
Eq.\,(\ref{eq:GF_0}) we see that the dot Majorana fermions hybridize with the
$a(0)$ field, leaving the $b(0)$ field decoupled. In the special case 
$\epsilon_{\textrm{d}}=0$, while the $\gamma_2$ mode still couples to the $a(0)$ field, 
the $\gamma_1$ mode is now totally decoupled [see Eqs.\,(\ref{eq:G0gamgam}) and (\ref{eq:G0agam})]. For $\lambda=0$, this corresponds
to the Majorana quantum critical state described by the solvable Emery-Kivelson point already discussed in Sec.~\ref{Transport}.

\subsection{General conductance formula}

Because the $b(0)$ field does not couple to any other Majorana modes, and
since the contact interaction Eq.\,(\ref{eq:HTHC_MR}) does not involve $b(0)$ 
either, the Green function of $b(0)$ can be exactly separated out in the 
current-current correlator of Eq.\,(\ref{eq:CII}), even in the interacting case
$\lambda\neq0$. 
It readily follows that the linear-response conductance can be written in terms 
of only the full spectral function of the $\gamma_2$ Majorana fermion, given by 
\begin{equation}
 A_{\gamma_2}(\omega)=-\text{Im}G^{R}_{\gamma_2\gamma_2}(\omega). 
\label{eq:SpectralFunction}
\end{equation}
The prefactor of the conductance is fixed by taking into account the Fermi-liquid nature of electrons in the source and drain reserviors; thus, the maximum conductance is $e^2/h$ instead of $ge^2/h$ \cite{MaslovPRB95,SafiSchulzPRB95,PonomarenkoPRB95}.
We thus find that
\begin{equation}
 G=\frac{e^2}{h}\int d\omega \Gamma A_{\gamma_2}(\omega)
\left( -\frac{\partial n_F(\omega)}{\partial \omega}\right),
\label{eq:G_SF}
\end{equation}
where $n_F(\omega)$ is the Fermi distribution function.
\emph{This is one of the main results of the paper: it shows that the
interacting Majorana transport theory can be formulated within a simple
Landauer-type expression involving the full Majorana spectral function.} This
expression is similar to the well-known Meir-Wingreen formula for the
conductance through an interacting quantum dot \cite{MeirWingreenPRL92}. Indeed,
the conductance can usually be expressed this way when the leads are
non-interacting, which is not the case in our present study due to strong
dissipation in the leads. We note that a similar though more complicated
expression holds in the case of asymmetric coupling, $V_S \neq V_D$.  

At the Emery-Kivelson point $\lambda=0$, using Eq.\,(\ref{eq:GF_0}a), one
obtains an exact expression for the dimensionless conductance in the absence of  
contact interaction, as found previously by Komnik and Gogolin \cite{KomnikPRL03}:
\begin{equation}
 g_0=\frac{G_{\lambda=0}}{e^2/h}=\int d\omega 
\frac{\Gamma^2\omega^2}{(\omega^2-\epsilon_{\textrm{d}}^2)^2+\Gamma^2\omega^2} 
\left( - \frac{\partial n_F(\omega)}{\partial \omega}\right).
\label{eq:g0}
\end{equation}
In this equation, the structure of the spectral function is quite different 
from the familiar Lorentzian lineshape for resonant fermionic tunneling, because of 
the non-trivial effect of dissipation in the leads. 
At zero temperature, this Emery-Kivelson solution displays a quantum phase transition 
controlled by the detuning $\epsilon_{\textrm{d}}$ \cite{KanePRB92,FurusakiPRB93,EggertAffleckPRB92,MebrahtuNat12}: 
when $\epsilon_{\textrm{d}}=0$, the ground state is a conducting state with a unitary 
conductance $g_0(T=0)=e^2/h$, otherwise the conductance vanishes. 
We are mainly interested in the scaling behavior close to and away from 
the Majorana quantum critical point, in the presence of the contact interaction.

\begin{figure}[t!]
\centering
\includegraphics[width=0.35\textwidth]{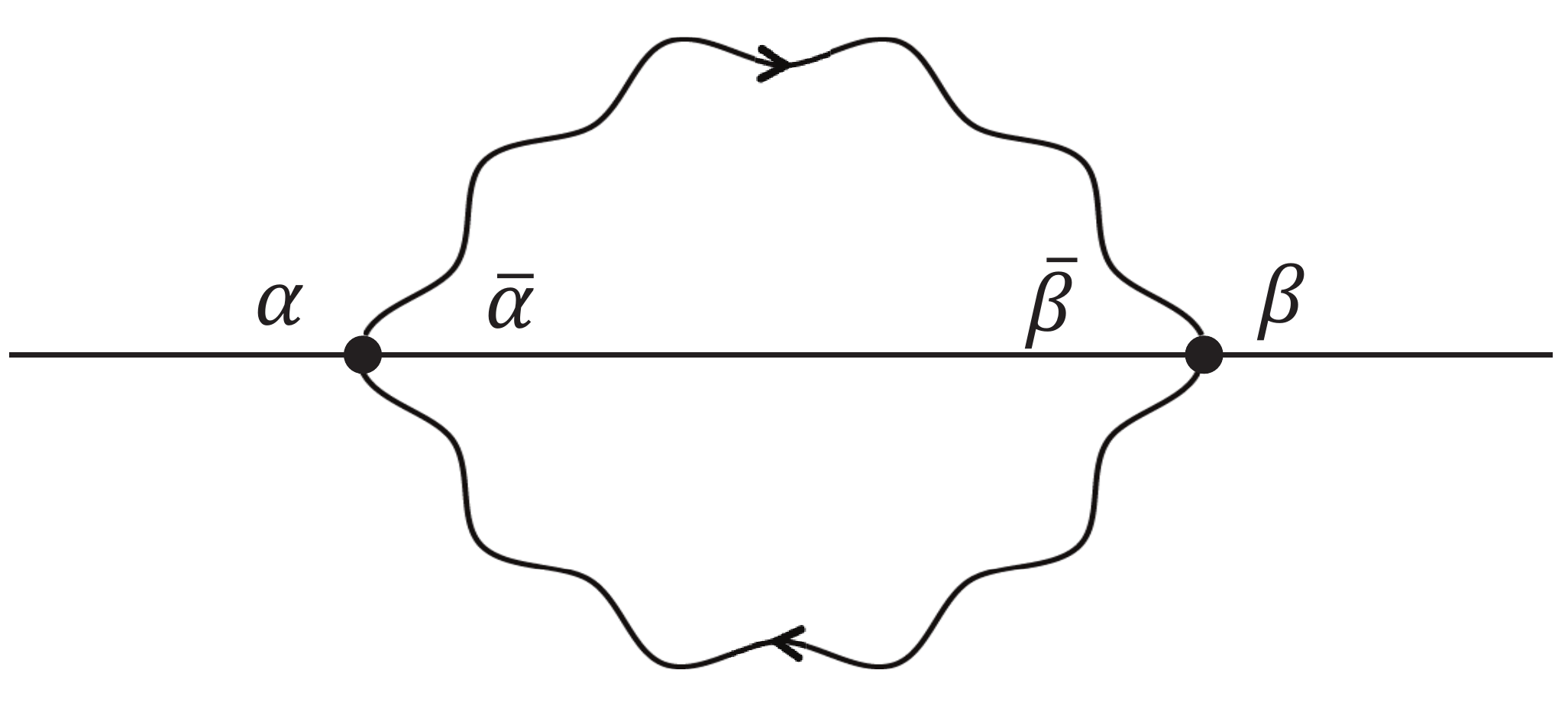}
\caption{\textbf{Second-order diagram of the resonant level Majorana fermion
self-energy.}  The bath $\langle \psi^{\dagger}_c\psi_c\rangle$ and 
Majorana $\langle \gamma_\alpha \gamma_\alpha \rangle$  propagators are represented 
by wiggly and straight lines, respectively.  Here, $\alpha=1,2$ label the two 
Majorana species, and we defined $\bar{\alpha}=1$ if $\alpha=2$ (and vice-versa).
}
\label{fig:2nd-diagram}
\end{figure}

\subsection{Perturbative treatment around the Emery-Kivelson point}
\label{Pert}

We now present perturbative results for the conductance away from 
the Emery-Kivelson point at order $\lambda^2$. A similar strategy was used 
previously to find thermodynamic quantities and the bulk resistivity in the two-channel Kondo 
context~\cite{SenguptaPRB94,ColemanIoffeTsvelik,ZitkoPRB11}.  
Straighforward calculations (see Appendix~\ref{App1}) give the following correction to the $\gamma_2$ propagator:
\begin{equation}
\label{PropFinal}
\delta G_{\gamma_2\gamma_2}^{(2)}(\omega)=\lambda^2\sum_{\alpha,\beta=1,2}(-1)^{\alpha+\beta}  
G^{(0)}_{\gamma_2\gamma_{\alpha}}(\omega) \Sigma^R_{\bar{\alpha}\bar{\beta}}(\omega) 
G^{(0)}_{\gamma_{\beta}\gamma_2}(\omega) ,
\end{equation}
where $\bar{\alpha}=1$ if $\alpha=2$ and vice-versa. The associated
self-energy matrix 
(see the diagram in Fig.\,\ref{fig:2nd-diagram}) 
reads
\begin{eqnarray}
\label{SelfFinal}
\Sigma^R_{\alpha \beta}(\omega)&=& \int 
\frac{d\omega_1d\omega_2}{\pi}\frac{(-\pi\rho^2\omega_1) 
\text{Im}[G^{(0)}_{\gamma_{\alpha}\gamma_{\beta}}(\omega_2) ]}{\omega+i\eta-\omega_1-\omega_2}\\
&& \nonumber \times \left[n_B(\omega_1)+n_F(-\omega_2) \right].
\end{eqnarray}
The resulting (dimensionless) second-order correction to the linear-response
conductance is therefore given by
\begin{equation}
\delta g_2=\frac{\delta G_2}{e^2/h}=\int d\omega \Gamma \delta A^{(2)}_{\gamma_2}(\omega)
\left(-\frac{\partial n_F(\omega)}{\partial \omega}\right),
\label{eq:corr_G}
\end{equation}
where the second-order correction to the spectral density is
$\delta A^{(2)}_{\gamma_2}(\omega)=-\text{Im}[\delta G_{\gamma_2\gamma_2}^{(2)}(\omega)]$.
Eqs.\,(\ref{eq:g0}-\ref{eq:corr_G}) are the central results of this paper;
they allow us to investigate the various scaling laws related to dissipative tunneling.

\section{Analysis of the transport scaling laws}
\label{Scaling}

In this section, we study in detail the scaling laws, and examine three
different regimes: (i) large detuning (Sec.~\ref{sec:large}); (ii) perfect
tuning at the Majorana quantum critical point (Sec.~\ref{sec:perfect}); (iii) small
detuning away from the quantum critical point (Sec.~\ref{sec:small}).
The main question to be addressed is whether the scaling laws 
derived from the non-interacting Hamiltonian at the Emery-Kivelson point are
modified by the perturbation of the contact interaction.

\begin{figure}[tb]
\centering
\includegraphics[width=0.45\textwidth]{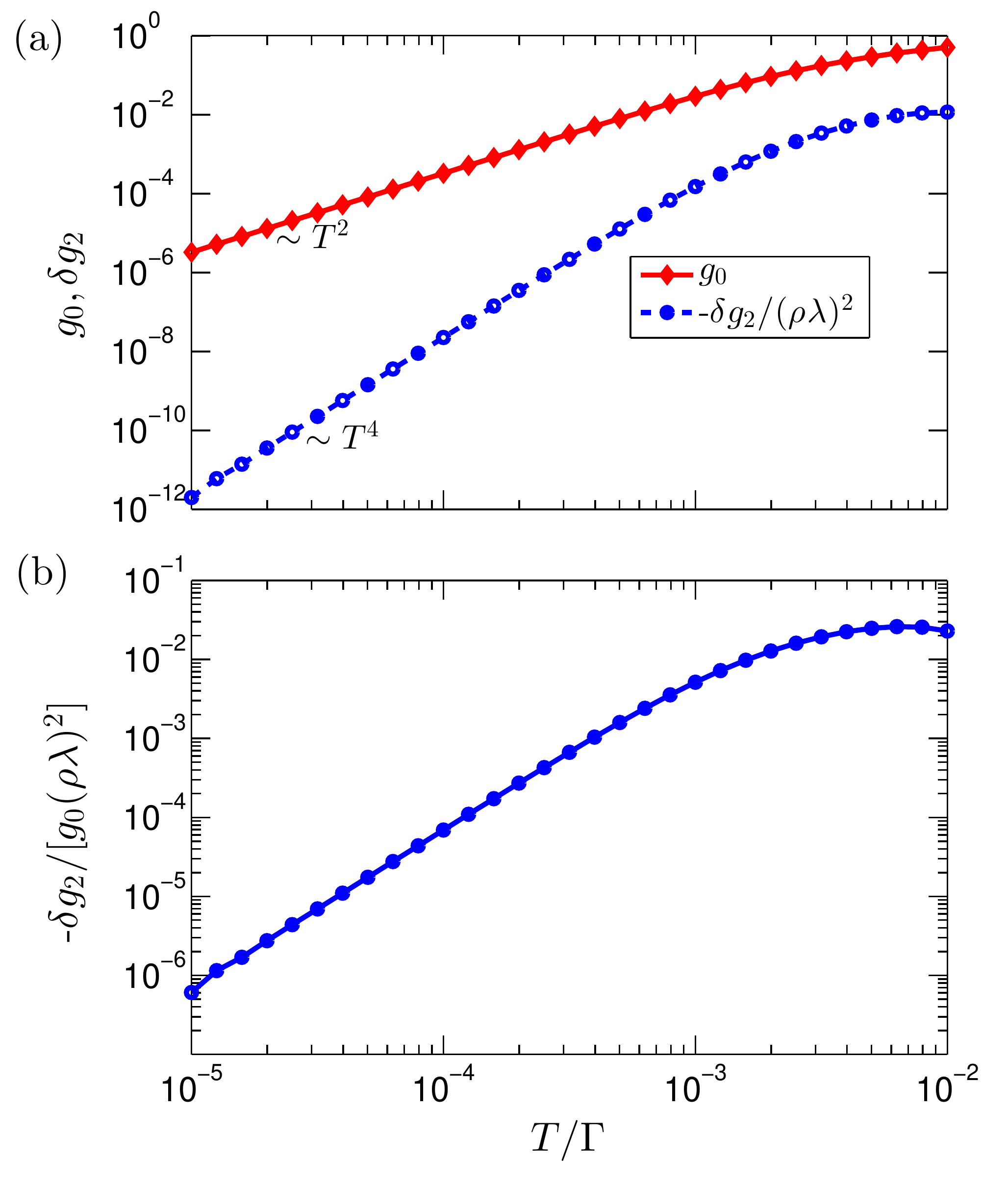}
\caption{(color online) \textbf{Large detuning, conductance shows single-barrier scaling.}
(a)~Low-temperature behavior of the conductance $g_0$ at
the Emery-Kivelson point (red, diamond) and the interaction-driven correction
$\delta g_2$ (blue, circle), in the regime of sizeable detuning 
(here $\epsilon_{\textrm{d}}=0.1\,\Gamma$ 
for which $\Gamma' \approx 0.02 \,\Gamma$), 
as a function of $T/\Gamma$.
(b)~The small dimensionless ratio $-\delta g_2/[g_0(\rho \lambda)^2]$ indicates
the validity of the single-barrier scaling in the present case.}
\label{fig:SB}
\end{figure}

\subsection{Large detuning: single barrier scaling}
\label{sec:large}

The simplest situation is that of a deep level in the quantum dot,
$|\epsilon_{\textrm{d}}| \gtrsim \Gamma'$, where $\Gamma'$ is the low-energy
renormalized width of the resonance (which can be much smaller than $\Gamma$).
As a result, the electrons tunnel through the system in a single process 
(co-tunneling)~\cite{IngoldNazarov92},
with only virtual occupation of the resonant level.
In this case, the backscattering operator (in the bosonization formulation) 
is relevant at low temperatures. The backscattering drives the system to an insulating 
state~\cite{KanePRB92,FurusakiPRB93,EggertAffleckPRB92,Sassetti_Napoli_Weiss_95,PolyakovPRB03,BomzePRB09}.  
Thus, the exact solution $g_0(T)$ at
the Emery-Kivelson point in this situation should have the same low-temperature
scaling as the conductance in tunneling through a strong single barrier \cite{KanePRB92} 
in a Luttinger liquid, namely $g_0(T)\propto T^{2(1/g-1)}=T^2$ at low temperature. 
This was indeed verified in Ref.\,\onlinecite{KomnikPRL03} and can be seen 
by performing the integral in Eq.\,(\ref{eq:g0}) at $T\rightarrow 0$ for large
detuning,
\begin{equation}
g_0\approx \int d\omega \frac{\Gamma^2\omega^2}{\epsilon_{\textrm{d}}^4}
\frac{\beta \text{e}^{\beta\omega}}{(1+\text{e}^{\beta\omega})^2}
=\frac{\pi^2}{3}\left(\frac{T}{\Gamma}\right)^2\left(\frac{\Gamma}{\epsilon_{\textrm{d}}}\right)^4.
\end{equation}


The contact interaction should, for small $\lambda$, become ineffective in this
limit: when the dot dynamics is frozen, the contribution of the contact
interaction to $\delta g_2$ is irrelevant. Analyzing the asymptotic
low-temperature scaling of Eq.\,(\ref{eq:corr_G}), we find indeed
\begin{equation}
\delta g_2\propto -\left(\frac{T}{\Gamma}\right)^4\left(\frac{\Gamma}{\epsilon_{\textrm{d}}}\right)^6.
\end{equation}

Figure\,\ref{fig:SB}(a) shows the results for $g_0$ and $\delta g_2$ at
$\epsilon_{\textrm{d}}=0.1\,\Gamma$ after performing the numerical integrals in
Eqs.\,(\ref{eq:g0}) and (\ref{eq:corr_G}). Although $\epsilon_{\textrm{d}}$ is
not very large for this particular example, the single-barrier scaling law
is already remarkably well obeyed.
The observed low-temperature scaling ($\sim T^2$ for $g_0$ and $\sim T^4$ for 
$\delta g_2$) confirms our asymptotic analysis.

Figure\,\ref{fig:SB}(b) plots the ratio between $\delta g_2$ and $g_0$
normalized by the dimensionless perturbation parameter $(\rho\lambda)^2$, which
should be less than $1$ to validate the perturbation theory. In the
low-temperature regime, this ratio is much smaller than $1$ and scales to zero
as $T^2$. Therefore, we conclude that including the contact interaction
term perturbatively up to second-order does not modify the
low-temperature single barrier scaling at the insulating fixed
point. This finding corroborates the experimental observation
\cite{MebrahtuNatPhys13} of the applicability of single-barrier scaling 
\cite{AristovEPL08} to describe the dissipative resonant-level system away
from the resonance.

\subsection{Low-temperature scaling at the conducting critical point}
\label{sec:perfect}

We now consider the case of perfect tuning to the quantum critical
point $\epsilon_{\textrm{d}}=0$, and focus on the low-temperature approach
to the unitary conductance~\cite{MebrahtuNat12} for $T\ll\Gamma$. 
By solving for the exact solution at the Emery-Kivelson point ($\lambda=0$), 
Komnik and Gogolin~\cite{KomnikPRL03} pointed out that the approach obeys a Fermi liquid 
form~\cite{PolyakovPRB03}, as can be checked in the considered regime 
from Eq.~(\ref{eq:g0}):
\begin{equation}
g_0= 1- \int d\omega\frac{\omega^4}{\omega^4+\Gamma^2\omega^2}
\frac{\beta \text{e}^{\beta\omega}}{(1+\text{e}^{\beta\omega})^2}\approx 
1- \frac{\pi^2}{3}\left(\frac{T}{\Gamma}\right)^2.
\end{equation}
This result however corresponds to an exact and unfortunate cancellation of 
the leading irrelevant operator~\cite{EmeryKivelsonPRB92,SenguptaPRB94} at the conducting
fixed point.

   From Eq.\,(\ref{eq:GF_0}), we observe that when $\epsilon_{\textrm{d}}=0$ only half of the
Majorana modes (namely $\gamma_2$) hybrize with the leads, leaving the $\gamma_1$
Majorana fermion fully decoupled from the rest of the system. Including the
contact interaction term $\lambda$ does not destroy the isolated Majorana mode;
however, it does give rise to an anomalous non-Fermi liquid temperature dependence.  
In the resonant case, because the Green function between $\gamma_1$
and $\gamma_2$ vanishes [see Eq.\,(\ref{eq:GF_0}a)], the only non-zero
correction to the $\gamma_2$ propagator in 
Fig.\,\ref{fig:2nd-diagram} is
\begin{equation}
\label{eq:g2_delta0}
\delta g_2=\int d\omega \Gamma \left(-\text{Im}\left[\lambda^2  
\left(G^{(0)}_{\gamma_2\gamma_2}(\omega)  \right)^2 \Sigma^R_{11}(\omega) \right] \right) 
\left(-\frac{\partial n_F(\omega)}{\partial \omega}\right).
\end{equation}
For $\epsilon_{\textrm{d}}=0$, $G^{(0)}_{\gamma_2\gamma_2}(\omega) =1/(\omega+i\Gamma)$ and 
$G^{(0)}_{\gamma_1\gamma_1}(\omega) =1/(\omega+i\eta)$. 
Hence, $\text{Im}[G^{(0)}_{\gamma_1\gamma_1}(\omega) ]=-\pi \delta(\omega)$. 
The self-energy $\Sigma^R_{11}$ can be evaluated readily
\begin{subequations}
\label{eq:Sigma_11}
\begin{eqnarray}
\Sigma^R_{11}(\omega,\;\epsilon_{\textrm{d}}=0)&=&\frac{\rho^2}{\beta}[P_1(\beta\omega)+iP_2(\beta\omega)], \\
P_1(\beta\omega)&=&\fint dx \frac{x}{\beta\omega-x}\frac{1}{2}\text{coth}\left(\frac{x}{2}\right), \\
P_2(\beta\omega)&=&-\frac{\pi\beta\omega}{2}\text{coth}\left(\frac{\beta\omega}{2}\right).
\end{eqnarray}
\end{subequations}
Plugging Eq.\,(\ref{eq:Sigma_11}) into Eq.\,(\ref{eq:g2_delta0}), we have
\begin{eqnarray}
\label{eq:g2_d0}
&& \delta g_2=(\rho\lambda)^2\int d\omega\left( -\frac{\Gamma}{\beta} \right) 
\Bigg[ \frac{\omega^2-\Gamma^2}{(\omega^2+\Gamma^2)^2}P_2(\beta\omega) 
\qquad\qquad\qquad \nonumber \\
&& \qquad\qquad\qquad\qquad\quad 
-\frac{2\omega\Gamma}{(\omega^2+\Gamma^2)^2}P_1(\beta\omega)  \Bigg] 
\frac{\beta \text{e}^{\beta\omega}}{(1+\text{e}^{\beta\omega})^2}. 
\end{eqnarray}
In the low-temperature limit, the $P_2$ part dominates, and we 
obtain the following asymptotic scaling for $T\ll \Gamma$:
\begin{equation}
\label{eq:g2-T}
\frac{\delta g_2}{(\rho\lambda)^2}\approx 
\int d\omega\left( -\frac{\Gamma}{\beta} \right)\frac{1}{\Gamma^2} 
\frac{\pi\beta\omega}{2}\text{coth}\left(\frac{\beta\omega}{2}\right) 
\frac{\beta \text{e}^{\beta\omega}}{(1+\text{e}^{\beta\omega})^2}  
=-\frac{\pi^3}{8} \frac{T}{\Gamma}.
\end{equation}

This striking $T$ dependence is a strong signature of the uncoupled Majorana 
mode $\gamma_1$. Indeed, on resonance $\epsilon_{\textrm{d}}=0$, the correlation 
function of $\gamma_1$ does not decay at long time, 
$G_{\gamma_1\gamma_1}(t)=-\langle \gamma_1(0)\gamma_1(t)\rangle \propto 1$, instead
of the usual $1/t$ decay for hybridized modes. This translates into a $1/t^2$
decay of the $\gamma_2$ self-energy correction (instead of $1/t^3$ for a usual
Fermi liquid), giving rise by Fourier transform to a linear in frequency scattering 
rate.
This linear approach to the unitary conductance signals the presence
of an isolated Majorana state~\cite{SenguptaPRB94,ColemanIoffeTsvelik,ZitkoPRB11}, and 
has been observed in a recent experiment \cite{MebrahtuNatPhys13}. 
 
\begin{figure}[tb]
\centering
\includegraphics[width=0.45\textwidth]{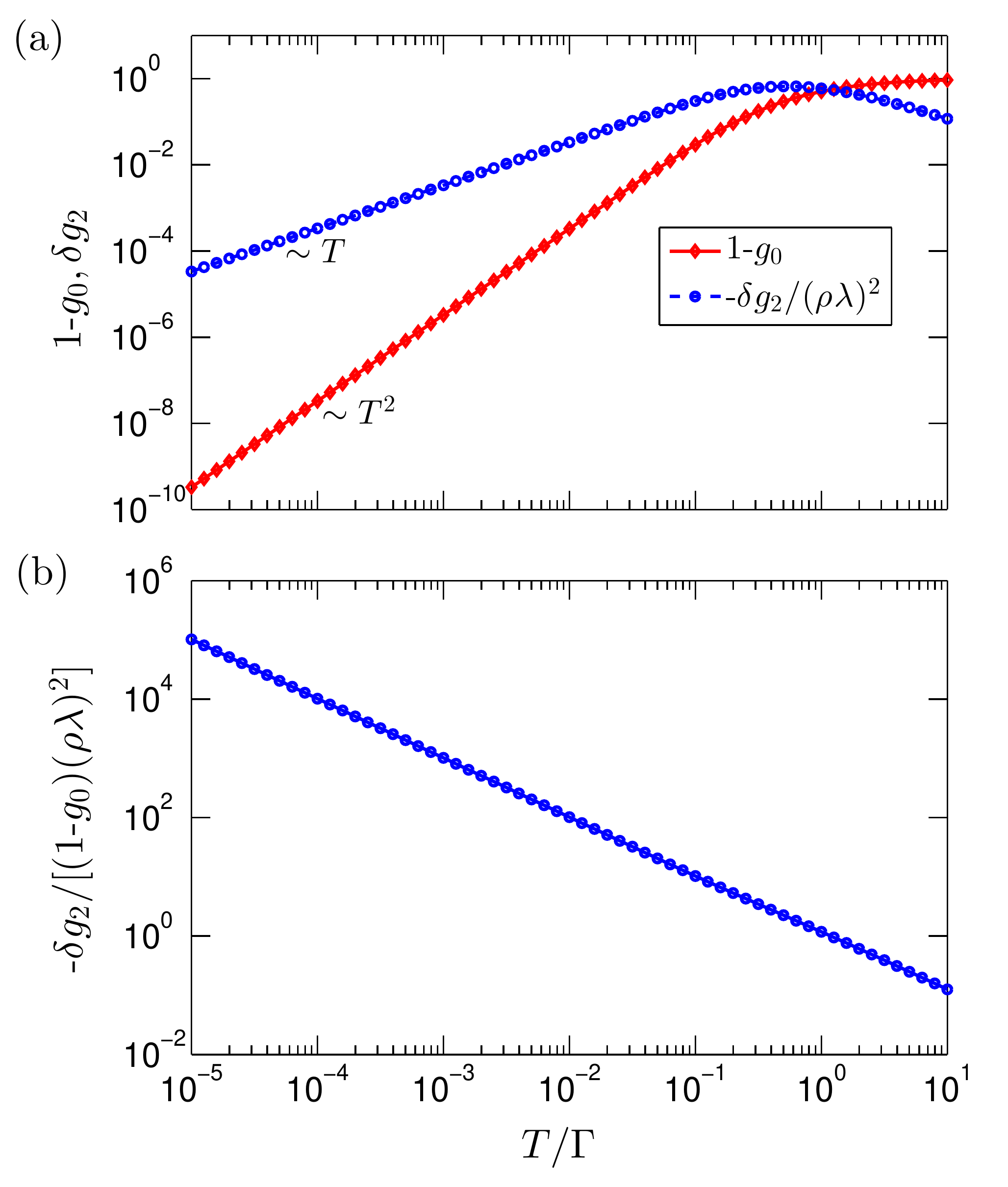}
\caption{(color online) \textbf{Approach to the quantum critical point; here $\epsilon_{\textrm{d}}=0$.}
(a)~Low-temperature behavior of $1-g_0$ (red, diamond)
and $\delta g_2$ (blue, circle) close to the conducting quantum critical point. 
(b) The scaling of $-\delta g_2/[(1-g_0)(\rho \lambda)^2]$ as a function of $T/\Gamma$ shows that the interaction correction dominates.
}
\label{fig:Runaway}
\end{figure}

Figure\,\ref{fig:Runaway}(a) shows both $1\!-\!g_0$ and $\delta g_2$ obtained by
numerical integration.  The asymptotic scalings are reproduced at low
temperatures.  Figure\,\ref{fig:Runaway}(b) plots the ratio of 
$-\delta g_2$ to $1-g_0$ normalized by the dimensionless perturbation parameter 
$(\rho \lambda)^2$.  As long as $(\rho\lambda)^2$ is not too small, the linear
temperature scaling strongly dominates over the quadratic
behavior as $T\!\rightarrow\! 0$. Hence, we conclude that the contact interaction
between the Majorana modes and the effective leads generates non-Fermi liquid
behavior at the Majorana quantum critical point.

Note that the four-fermion interaction term in Eq.~(\ref{eq:HMajorana}) is too
large ($\lambda = \pi v_F$) for the perturbation theory to quantitatively
capture the full crossover from high temperature ($T\gg \Gamma'$) to the asymptotic
non-Fermi liquid regime ($T\ll \Gamma'$), where $\Gamma'\ll\Gamma$ is the strongly
renormalized linewidth. 
This strong coupling regime~\cite{ZitkoSimonPRB11} leads
to universal scaling relations describing the full crossover towards the quantum
critical state in our system.

\begin{figure}[tb]
\centering
\includegraphics[width=0.45\textwidth]{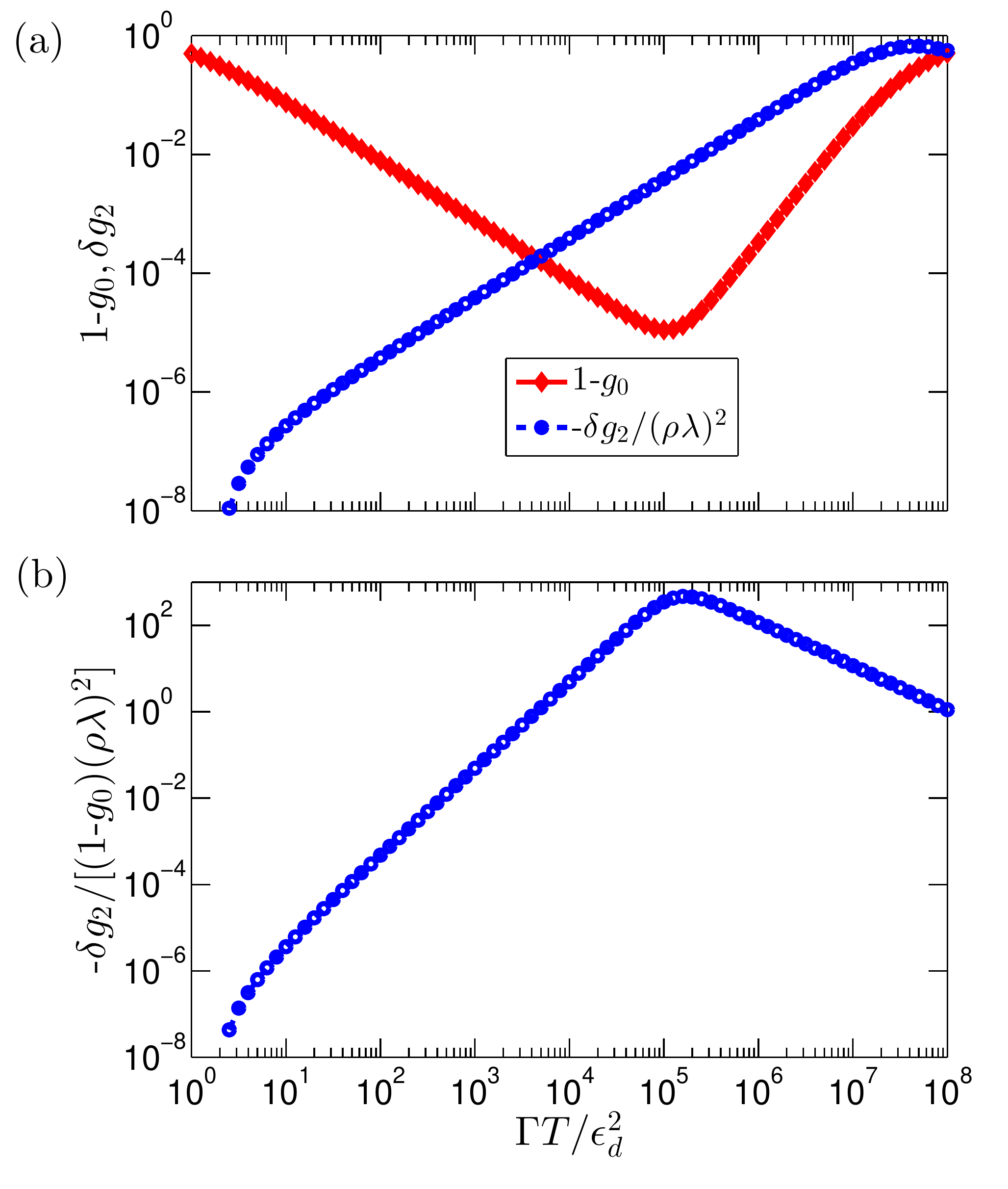}
\caption{(color online) \textbf{Small detuning, the runaway flow.}
Here, we choose $\epsilon_{\textrm{d}}=10^{-4}\Gamma$.
(a) $1-g_0$ (red, diamond) and $\delta g_2$ (blue,
circle) as a function of $\Gamma T/\epsilon_{\textrm{d}}^2$. 
(b) The ratio $-\delta g_2/[(1-g_0)(\rho \lambda)^2]$ as a function of $\Gamma
T/\epsilon_{\textrm{d}}^2$. Initially, the interaction corrections
dominate as one approaches the critical point, but then the system veers away
toward the insulating fixed point and the non-interacting term, $g_0$, dominates
in the end.  
}
\label{fig:11SB}
\end{figure}

\subsection{Small detuning: runaway flow}
\label{sec:small}

We finally investigate intermediate-temperature scaling with
a slight detuning from the quantum critical point,
$T \ll \epsilon_d \ll \sqrt{T\Gamma}$.
In that regime, the renormalization flow approaches very close to the
conducting fixed point, but ultimately flows away from it because the transparency
is not perfectly unity. Considering first the Emery-Kivelson solution
Eq.\,(\ref{eq:g0}) in this limit, we obtain the runaway behavior from the unitary 
conductance, which has the same $1/T$ temperature dependence as tunneling 
through a weak single barrier \cite{KanePRB92, KomnikPRL03}: 
\begin{eqnarray}
\label{eq:g0_C}
g_0&=&1-\int d\omega\frac{(\omega^2-\epsilon_{\textrm{d}}^2)^2}{(\omega^2-\epsilon_{\textrm{d}}^2)^2+\Gamma^2\omega^2}
\frac{\beta \text{e}^{\beta\omega}}{(1+\text{e}^{\beta\omega})^2} \qquad\qquad \nonumber \\
&\approx 1- & \int d\omega \frac{(\epsilon_{\textrm{d}}^2/\Gamma)^2}{\omega^2+(\epsilon_{\textrm{d}}^2/\Gamma)^2} 
\frac{\beta \text{e}^{\beta\omega}}{(1+\text{e}^{\beta\omega})^2} \nonumber \\
&\approx 1-& \int d\omega \frac{(\epsilon_{\textrm{d}}^2/\Gamma)^2}{\omega^2+(\epsilon_{\textrm{d}}^2/\Gamma)^2} 
\frac{\beta \text{e}^{0}}{(1+\text{e}^{0})^2}
= 1- \frac{\pi}{4} 
\left( \frac{\epsilon_{\textrm{d}}^2}{\Gamma T} \right).
\end{eqnarray}
In Eq.\,(\ref{eq:g0_C}), we used in the second and third lines the conditions 
$\Gamma \gg \epsilon_{\textrm{d}} \gg T$ and $T \gg \epsilon_{\textrm{d}}^2/\Gamma$,
respectively.

On the other hand, $\delta g_2$ still obeys Eq.\,(\ref{eq:g2-T}), since $\Gamma\gg
\epsilon_{\textrm{d}}$, $T$.  Therefore, we have the ratio
\begin{equation}
 -\frac{\delta g_2}{1-g_0}\approx
\frac{\pi^2}{2}(\rho\lambda)^2\left(\frac{T}{\epsilon_{\textrm{d}}} \right)^2,
\end{equation}
which is much smaller than $1$ for $T\ll \epsilon_{\textrm{d}}$, indicating that the
runaway flow of $1-g_0$ is not modified by the perturbation correction from the
contact interaction term.

Figure\,\ref{fig:11SB}(a) presents $1-g_0$ and $\delta g_2$ as a function of
$\Gamma T/\epsilon_{\textrm{d}}^2$ with a small detuning $\epsilon_{\textrm{d}}=10^{-4}\Gamma$ over a wide
temperature range.  For very low temperature $T\sim \epsilon_{\textrm{d}}^2/\Gamma$, $1-g_0\sim
1$ showing that even a small detuning can drive the system to the insulating
critical point with a vanishing conductance.  In the intermediate-temperature
regime ($10^1\lesssim \Gamma T/\epsilon_{\textrm{d}}^2\lesssim 10^3$), the condition $\Gamma\gg
\epsilon_{\textrm{d}}\gg T \gg \epsilon_{\textrm{d}}^2/\Gamma$ is satisfied.  Clearly, Fig.\,\ref{fig:11SB}(b)
shows that in this temperature range $\delta g_2$ is subdominant compared to
$1-g_0$.  Further increase of temperature leads to the regime $\Gamma\gg
T\gtrsim \epsilon_{\textrm{d}}$ ($10^4\lesssim \Gamma T/\epsilon_{\textrm{d}}^2\lesssim 10^7$).  In this
regime, $1-g_0$ changes from $1/T$ to $T^2$ dependence and $\delta g_2$ starts
to dominate the runaway scaling as shown in Fig.\,\ref{fig:11SB}(b).

To conclude this study, we give in Table\,\ref{tb:scalings} a summary of the scalings 
in the three different regimes discussed in this section. The contact
interaction controls the approach to the quantum critical point, but is strongly
irrelevant otherwise, leading to effectively single barrier scaling.

\begin{table}[tb]
\renewcommand{\arraystretch}{2.5}   
\setlength{\tabcolsep}{3pt}         
\begin{tabular}{|c||c|c|c|}
\hline 
\text{Regime}  &  $g_0$  & $1-g_0$  & $-\delta g_2/(\rho\lambda)^2$ \\
\hline
$\epsilon_{\textrm{d}}\sim \Gamma\gg T$  &  $\displaystyle \sim \frac{\pi^2}{3}\left(\frac{T}{\Gamma}\right)^2 \left(\frac{\Gamma}{\epsilon_{\textrm{d}}}\right)^4$ & $\sim 1$ & $\displaystyle \propto \left(\frac{T}{\Gamma}\right)^4 \left(\frac{\Gamma}{\epsilon_{\textrm{d}}}\right)^6$ \\[1.5mm]
\hline
$\epsilon_{\textrm{d}}=0$, $\Gamma \gg T$  & $\sim 1$ & $\displaystyle \sim \frac{\pi^2}{3}\left(\frac{T}{\Gamma}\right)^2$ & $\displaystyle \sim \frac{\pi^3}{8}\left(\frac{T}{\Gamma}\right)$ \\[1.5mm]
\hline
$\displaystyle \Gamma\gg\epsilon_{\textrm{d}}\gg T\gg \frac{\epsilon_{\textrm{d}}^2}{\Gamma}$ & $\sim 1$  & $\displaystyle \sim \frac{\pi}{4}\left( \frac{\epsilon_{\textrm{d}}^2}{\Gamma T}\right)$ & $\displaystyle \sim \frac{\pi^3}{8}\left(\frac{T}{\Gamma}\right)$ \\[1.5mm]
\hline
\end{tabular}
\caption{\textbf{Summary of various low-temperature scalings close to the insulating and
conducting fixed points.} The first, second, and third rows correspond to large detuning, exactly critical tuning, and small detuning (runaway flow), respectively.}
\label{tb:scalings}
\end{table}

\section{Conclusion}
In summary, we have studied spinless resonant tunneling with a large, fine-tuned circuit impedance $R=e^2/h$ and mapped it directly to resonant tunneling
between Luttinger liquids with Luttinger parameter $g=1/2$. We further mapped
the system to a resonant Majorana model in the case of symmetric coupling. In
contrast to previous studies, we retained the contact interaction between the
resonant level and the leads. Perturbation theory of the linear-response
conductance is developed up to second-order in the contact interaction. We found
that while the second-order correction does not change the single-barrier
scaling near the insulating fixed point, it does give rise to a linear
temperature dependence as the conductance approaches unity when the resonant
level is tuned to be exactly on resonance (Majorana quantum critical
point). This striking non-Fermi liquid
behavior is due to the fact that the resonant level is fractionalized into two
independent Majorana fermions, with one of them fully isolated from the rest of
the system. Further investigations could, for instance, concentrate on
incorporating the spin degree of freedom on the quantum dot, leading to a rich
interplay of Luttinger and Kondo physics.

\section*{Acknowledgments}

We thank G. Finkelstein for motivating this study. We acknowledge funding from the Fondation Nanosciences de Grenoble under RTRA contract CORTRANO, and from the US DOE, Division of Materials Sciences and Engineering, under Grant No.\,{DE-SC0005237}.

\appendix

\section{Derivation of the self-energy correction}
\label{App1}

The diagrammatic calculations proceed by expanding the propagator of
the $\gamma_2$ Majorana mode in powers of the 
contact interaction term $H_C$:
\begin{subequations}
\label{eq:Gd2d2_Pert}
\begin{eqnarray}
 G_{\gamma_2\gamma_2}(\tau)&=&-\langle T_{\tau} \gamma_2(\tau)\gamma_2(0)\rangle
\\
\nonumber
&=& -\Big\langle T_{\tau} \Big\{ \gamma_2(\tau)\gamma_2(0)
\text{exp}\Big[-\int^{\beta}_{0}d\tau H_C(\tau) \Big] \Big\} \Big\rangle_0\\
\nonumber
&=& -\sum_{n} \Big \langle   T_{\tau} \Big\{ \gamma_2(\tau)\gamma_2(0)\frac{1}{n!} 
\Big[ -{\textstyle \int} d\tau^{\prime}H_C(\tau^{\prime})\Big]^n \Big\} \Big \rangle_0. 
\end{eqnarray}
\end{subequations}
The zeroth-order contribution provides the non-interacting contribution already
given in the conductance formula (\ref{eq:g0}). 
The first-order contribution vanishes due to a disconnected diagram of the $\psi_c$ 
field under the non-interacting Hamiltonian. We therefore focus
on the second-order contribution, which gives rise to a correction to the
spectral function in Eq.\,(\ref{eq:SpectralFunction}) and hence to a correction
to the linear
response conductance.  The diagram for the second-order perturbation is shown in
Fig.\,{\ref{fig:2nd-diagram}} and reads
\begin{eqnarray}
\nonumber
\!\!\!\delta G_{\gamma_2\gamma_2}^{(2)}(\tau)&=&\frac{\lambda^2}{2}\!\!\iint
\!\!\!\! d\tau_1d\tau_2 
\Big \langle  T_{\tau} \Big[  \!\colon 
\psi^{\dagger}_c(\tau_1)\psi_c(\tau_1)\psi^{\dagger}_c(\tau_2)\psi_c(\tau_2)  
\colon  \!\Big]  \Big \rangle_0 \\
\nonumber 
&& \times
\Big \langle T_{\tau} \Big[ \gamma_2(\tau)\gamma_2(0)  
\gamma_1(\tau_1)\gamma_2(\tau_1) \gamma_1(\tau_2)\gamma_2(\tau_2) \Big]  
\Big \rangle_0 \\
 \label{eq:Gd2_2nd}
&=& \lambda^2\iint d\tau_1d\tau_2 
\sum_{\alpha,\beta=1,2}(-1)^{\alpha+\beta} 
G^{(0)}_{\gamma_2\gamma_{\alpha}}(\tau-\tau_1) \\
\nonumber
&&\times\Sigma_{\bar{\alpha}\bar{\beta}}(\tau_1-\tau_2) 
G{(0)}_{\gamma_{\beta}\gamma_2}(\tau_2) ,
\end{eqnarray}
where $\bar{\alpha}=1$ if $\alpha=2$ (and vice-versa).

The self-energy of Majorana fermions is defined as
\begin{subequations}
\label{eq:self-energy}
\begin{eqnarray}
 && \Sigma_{\alpha\beta}(\tau)=G^{(0)}_c(\tau)  G^{(0)}_c(-\tau)  G^{(0)}_{\gamma_{\alpha}\gamma_{\beta}}(\tau) , \\
 && G^{(0)}_c(\tau) =-\langle T_{\tau} \psi^{\dagger}_c(x=0,\tau)\psi_c(x=0,0)\rangle_0.
\end{eqnarray}
\end{subequations}
After Fourier transformation of Eqs.\,(\ref{eq:Gd2_2nd}) and (\ref{eq:self-energy}), we have
\begin{subequations}
\label{eq:Gd2d2-freq}
\begin{eqnarray}
\nonumber
\delta G_{\gamma_2\gamma_2}^{(2)}(i\omega_n)&=&\!
\lambda^2  \!\!\! \sum_{\alpha,\beta=1,2}\!\!\! (-1)^{\alpha+\beta}  
G^{(0)}_{\gamma_2\gamma_{\alpha}}(i\omega_n) \Sigma_{\bar{\alpha}\bar{\beta}}(i\omega_n) 
G^{(0)}_{\gamma_{\beta}\gamma_2}(i\omega_n) , \\
&&\\
\Sigma_{\alpha \beta}(i\omega_n)&=&\frac{1}{\beta} 
\sum_{ip_n} \,\chi(ip_n) \,G^{(0)}_{\gamma_{\alpha}\gamma_{\beta}}(i\omega_n-ip_n) , \\
\chi(ip_n)&=& \frac{1}{\beta} \sum_{iq_n} G^{(0)}_c(ip_n+iq_n) \,G^{(0)}_c(iq_n) .
\end{eqnarray}

\end{subequations}
Here, $p_n$ and $q_n$ are bosonic and fermionic Matsubara frequencies,
respectively. The Matsubara sum over $iq_n$ can be done easily, since
$G^{(0)}_c(iq_n)$ has a simple pole \cite{BruusBook,MahanBook}, so that:
\begin{eqnarray}
\label{eq:Chi}
  \chi(ip_n)&=& \int d\epsilon_1 d\epsilon_2
\frac{1}{\beta}\sum_{iq_n}\frac{\rho}{iq_n+\epsilon_1}\frac{\rho}{ip_n+iq_n+\epsilon_2}\\
\nonumber
& = & \int d\epsilon_1 d\epsilon_2
\frac{\rho^2}{ip_n+\epsilon_1-\epsilon_2}\left[n_F(\epsilon_1)-n_F(\epsilon_2)\right].
\end{eqnarray}
To evaluate the self-energy, we rely on the following identity of Matsubara
Green functions \cite{MahanBook}
\begin{equation}
 G(i\omega_n)=-\int \frac{d\epsilon}{\pi}\frac{\text{Im}[G^R(\epsilon)]}{i\omega_n-\epsilon}.
\end{equation}
Using this, Eq.\,(\ref{eq:Gd2d2-freq}b) can be written as
\begin{eqnarray}
\nonumber
\Sigma_{\alpha \beta}(i\omega_n) &=& \frac{1}{\beta} \sum_{ip_n}  
\frac{ \int d\omega_1 d\omega_2 } {\pi^2} 
\frac{\text{Im}[\chi^{R}(\omega_1)]}{ip_n-\omega_1} 
\frac{\text{Im}[G^{(0)}_{\gamma_{\alpha}\gamma_{\beta}}
(\omega_2)]}{i\omega_n-ip_n-\omega_2}\\
\label{eq:Sigma}
& =& \int \frac{d\omega_1d\omega_2}{\pi}
\frac{\text{Im}[\chi^{R}(\omega_1)] 
\text{Im}[G^{(0)}_{\gamma_{\alpha}\gamma_{\beta}}(\omega_2) ]}{i\omega_n-\omega_1-\omega_2}\\
\nonumber
&& \times \left[n_B(\omega_1)+n_F(-\omega_2) \right],
\end{eqnarray}
where $n_B$ is the Bose-Einstein distribution function. Again, the summation
over $ip_n$ is straightforward because the integrand has only two
simple poles at $\omega_1$ and $i\omega_n-\omega_2$ \cite{MahanBook}.
Performing an analytic continuation and evaluating the integral in
Eq.\,(\ref{eq:Chi}), we obtain $\text{Im}[\chi^R(\omega)]=-\pi \rho^2 \omega$ in
the wide band limit.  After analytic continuation of
Eq.\,(\ref{eq:Gd2d2-freq}a) and Eq.\,(\ref{eq:Sigma}), we arrive at Eqs.\,(\ref{PropFinal})-(\ref{SelfFinal}) quoted in the main text.

\end{document}